\documentclass[fleqn,twoside]{article}
\usepackage[headings]{espcrc2}

\readRCS
$Id: espcrc2.tex,v 1.2 2004/02/24 11:22:11 spepping Exp $
\usepackage{graphicx}
\usepackage[figuresright]{rotating}

\newcommand{\AmS}{{\protect\the\textfont2
  A\kern-.1667em\lower.5ex\hbox{M}\kern-.125emS}}

\newcommand{\bean}{\begin{eqnarray*}}
\newcommand{\eean}{\end{eqnarray*}}

\newcommand{\gapproxeq}{\lower
.7ex\hbox{$\;\stackrel{\textstyle >}{\sim}\;$}}
\newcommand{\lapproxeq}{\lower
.7ex\hbox{$\;\stackrel{\textstyle <}{\sim}\;$}}

\newcommand\lsim{\mathrel{\rlap{\lower4pt\hbox{\hskip1pt$\sim$}}
    \raise1pt\hbox{$<$}}}
\newcommand\gsim{\mathrel{\rlap{\lower4pt\hbox{\hskip1pt$\sim$}}
    \raise1pt\hbox{$>$}}}
\newcommand{\ba}{\begin{array}}
\newcommand{\ea}{\end{array}}
\newcommand{\nn}{\nonumber}

\newcommand{\be}{\begin{equation}}
\newcommand{\ee}{\end{equation}}
\newcommand{\bear}{\begin{eqnarray}}
\newcommand{\eear}{\end{eqnarray}}
\newcommand{\tab}{\hspace*{0.5cm}}

\newcommand{\ket}{\,\rangle}
\newcommand{\bra}{\langle \,}
\newcommand{\eqn}[1]{(\ref{#1})}
\newcommand{\cO}{{\cal O}}
\newcommand{\bel}[1]{\be\label{#1}}

\newcommand{\mB}{\mathcal{B}}

\newcommand{\re}{r_\epsilon^2 }

\newcommand{\Frac}[2]{\frac{\displaystyle #1}{\displaystyle #2}}

\title{
Effective Theory 
Description 
of Weak Annihilation in $B\to X_u \ell \nu$ Decays
}

\author{
J.J. Sanz-Cillero\address[JJSC]{
	Groupe Physique Th\'eorique, IPN Orsay, 
	Universit\'e Paris-Sud XI,  91406 Orsay, France
   		\\
	e-mail: cillero@ipno.in2p3.fr
   	}%
	\thanks{
		Thanks to the organisers of BEACH 2006, Lancaster, UK, 
		and to C. Smith, E. Gardi, M.Ciuchini  
		and L. Oliver for useful comments and suggestions. 
		Also thanks to M. Neubert for his help and contributions  
		in early stages of the project.  
		Work done in collaboration with G. Paz~\cite{preparation}.
		and supported by EU~RTN~Contract~CT2002-0311.
	}}
       
\begin{document}

\begin{abstract}
The semileptonic $B\to X_u \ell \nu$ decays  allow a pretty 
clean determination of  
the CKM matrix element $|V_{ub}|$. 
Nevertheless, the presence of
weak annihilation effects near the end-point 
of the $q^2$ spectrum 
introduces uncertainties in the inclusive calculation,
requiring  the use of
non-perturbative techniques like heavy meson chiral perturbation theory
and large $N_C$ limit.
\vspace{1pc}
\end{abstract}

\maketitle


\section{Introduction}

The $B_q\sim b\bar{q}$ 
meson decay through the inclusive semileptonic channel $B_q\to X_u \ell \nu$ 
has been shown to be a very clean way to determine $|V_{ub}|$~\cite{BLNP}.

The inclusive $B_q$ analysis makes use 
of the operator product expansion (OPE)~\cite{BLNP,Fazio}.  
The first difference between the $B_u^-$ and $B_{d,s}^0$ decays arises  
at next-to-next-to-next-to-leading order (N$^3$LO) in the $1/m_b$ expansion.  
It is due to the topologies where the incoming $b\bar{q}$
pair annihilates into a $W$--boson (Fig.~\eqn{fig.WAdiagram}),   
giving away an infinitesimal amount of  momentum 
through soft gluon radiation. 
These contributions are known as weak-annihilation (WA)~\cite{Bigi}. 
They  appear as a delta function at the end-point of the
spectrum $q^2=(p_{\ell}+p_{\nu})^2=m_b^2$.  
Their determination is rather relevant if the uncertainty on $|V_{ub}|$
is to be reduced up to a few per cent~\cite{BLNP}.

In order to better isolate the WA, the analysis is carried on 
within the chiral limit $m_q=0$.

\begin{figure}[t!]
\begin{center}
\includegraphics[width=4.cm,clip]{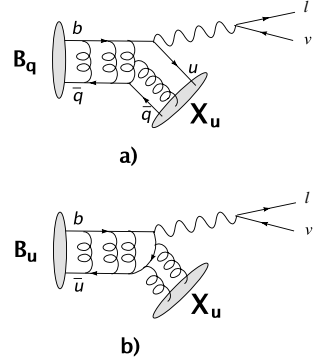}
\vspace*{-1cm}
\caption{
\small{
Examples of a) non-annihilating topologies; b) WA diagrams.}}
\vspace*{-1cm}
\label{fig.WAdiagram}
\end{center}
\end{figure}

\vspace*{-0.2cm}
\section{OPE and Weak-Annihilation}

At $\cO(1/m_b^3)$, the WA  in the OPE is  
given in terms  of the bag constants 
$B_{_{WA}}\equiv B_2-B_1$~\cite{Voloshin}:
\vspace*{-0.15cm}
\be
\Gamma(B_q\to X_u \ell\nu) = 
...\, +\,\Frac{G_F^2\, |V_{ub}|^2}{12\, \pi}\, 
f_B^2\, m_b^3\, \, B_{_{WA}}^{q}\, ,
\ee
with $B_{_{WA}}^{q}=0$ in the factorisation limit, independently of the 
light quark in the $B_q$ meson  ($B_2^{d,s}=B_1^{d,s}=1$ and 
$B_2^{u}=B_1^{u}=0$). 
The expected 10\% 
deviations from the factorisation limit provide an estimate 
of the WA contribution to the branching ratio (BR)~\cite{Voloshin}:
%
\vspace*{-0.2cm}
\be
\delta \mB_{B_q\to X_u \ell \nu}
= 3.9  \times  \left( \Frac{f_B}{0.2\mbox{ GeV}}\, |V_{ub}|\right)^2
\left(\Frac{B_{_{WA}}^q}{0.1}\right)\, .
\ee
Other theoretical and experimental estimates provide similar
values~\cite{Gambino,Meyer}.

The OPE calculation also points out 
relevant aspects about the nature and the scalings of the WA contributions:

\tab --Constant contribution to the width, generated at the end-point, 
in the non-perturbative region of low hadronic energies 
$E_X\lsim 1$~GeV~\cite{Bigi}.

\tab --The WA operators appear at N$^3$LO 
in the heavy mass expansion, 
so their contributions to the
BR is $\cO\left(\Lambda_{\rm QCD}^3/m_b^3\right)$.

\tab --The WA topologies introduce an
extra quark loop in the hadronic tensor $T^{\mu\nu}$, 
being suppressed by $1/N_C$ in the large $N_C$ limit~\cite{NC}. 

\tab --The WA $B^- - B^0$ difference is related 
to the production of chiral singlets in the final 
state $X_u$ (see Fig.~\eqn{fig.WAdiagram}). 
In the chiral limit, WA is 
the responsible of 
the difference between the $B^+$ and the $B^0$ widths. 

\tab --At N$^3$LO in $1/m_b$ there are 
other non-WA terms in the OPE (independent of $q$ in $B_q$) 
that need to be taken into account and properly disentangled~\cite{Gambino}.

\vspace*{-0.2cm}
\section{Cut-off on the hadronic energy $E_{X}$}

In order to isolate the WA effects and remove the $B\to X_c\ell\nu$
background  we consider the hadronic energy cut-off 
$E_{X}<\Lambda_{E}$.

Multi-pion production is found to be suppressed both experimentally~\cite{PDG} 
and theoretically.  
The analysis of the 
$B\to \pi\pi\ell\nu$ channel through Heavy Meson Chiral perturbation Theory 
(HM$\chi$PT)~\cite{hmchpt} 
finds it widely suppressed with respect to  $B\to\pi \ell \nu$ 
for $\Lambda_E\lsim 1$~GeV~\cite{preparation}.

Hence, only the lightest pseudo-scalar meson production 
$B\to P\ell\nu$ is considered here.  
The decay into light vector mesons $B\to V\ell\nu$ introduces 
small corrections for $\Lambda_E\lsim 1$~GeV and it 
can be neglected in first approximation~\cite{preparation}.

In the $B_q\to P \ell \nu$ mode, one has 
$P^2_{X}=m_P^2$ and
$E_{X}=E_P$, with the phase-space given by
\vspace*{-0.2cm}
\bear
\frac{1}{2}\left( z-\sqrt{z^2-4\, r^2}\right) 
\hspace*{-0.25cm}&\leq \bar{x}\leq &\hspace*{-0.25cm}
\frac{1}{2}\left( z+\sqrt{z^2-4\, r^2}\right) \, ,
\nn \\
2\, r&\leq  z \leq &\epsilon\,  \, ,
\label{eq.phasespace}
\eear
where $z=\frac{2 E_P}{M_B}$, $\bar{x}=1-\frac{2E_\ell}{M_B}$, 
$\epsilon=\frac{2 \Lambda_E}{M_B}$ and~${r=\frac{m_P}{M_B}}$.

The decay is determined by the matrix element 
\vspace*{-0.5cm}
\be
\bra P|\bar{u}\gamma^\mu (1-\gamma_5) b | B\ket 
= (p_B+p_P)^\mu \, f_+\, + \, q^\mu \, f_-\, ,
\ee
although, for $m_{\ell}=m_{\nu}=0$,  
only the form factor $f_+(z)$ contributes to the width:
\vspace*{-0.cm}
\be
\Frac{1}{\Gamma_0}\, \Frac{d\Gamma}{dz}\,\, =\,\, 
|f_+(z)|^2\, \, \, \cdot \, \, \, \left[z^2\, -\, 4 \, r^2\right]^\frac32 \, . 
\ee
with $\Gamma_0=\frac{G_F^2|V_{ub}|^2 M_B^5}{192\pi^3}$.  
Integrating the energy on the range $2r\leq z\leq
\epsilon$  provides the corresponding contribution to the width 
$\Delta \Gamma\left(\epsilon,r_\epsilon\right)$ from this part of the
phase-space. It only depends on $\epsilon$
and the ratio $r_\epsilon=\Frac{2\, r}{\epsilon}=\Frac{m}{\Lambda_E}$.

\vspace*{-0.2cm}
\section{HM$\chi$PT  and large $N_C$ limit}

To compute $f_+(z)$ and $\Delta \Gamma\left(\epsilon,r_\epsilon\right)$ 
we make use of HM$\chi$PT. At large $N_C$, the meson loops are suppressed. 
The $B\to P\ell\nu$ amplitude 
is dominated the $B^*$ pole from the tree-level exchange. 
Its description is provided by the leading HM$\chi$PT
lagrangian~\cite{hmchpt}: 
\vspace*{-0.5cm}
\bel{eq.FF}
\left| f_+(z)\right|_{B_q\to P}^2 \quad =\quad 
D_{B_q\to P}\quad \cdot\quad \Frac{f_B^2\, g^2}{2\, F_\pi^2}\quad
\Frac{1}{z^2}\, , 
\ee
with the group factors 
$D_{B_d\to \pi^+}=D_{B_s\to K^+}=1$, $D_{B_u\to \pi^0}=\frac{1}{2}$,  
$D_{B_u\to \eta_8}=\frac{1}{6}$,  $D_{B_u\to \eta_0}=\frac{1}{3}$. 
$f_B$ and $F_\pi\simeq 92.4$~MeV are the $B$ and $\pi$ 
decay constants, respectively, and $g$ is the $B\pi B^*$
coupling~\cite{hmchpt}.

In the large $N_C$ and chiral limits, the 
pseudoscalars $(\pi,K,\eta_8,\eta_0)$ 
are massless, so $r_\epsilon=0$.  
Therefore, the integrated widths coincide:
\vspace*{-0.3cm}
\bel{eq.LOFF}
\Delta \Gamma(\epsilon)_{B_d^0,\, B_s^0,\, B_u^+\to X_u \ell \nu}
\,= \, \Frac{f_B^2 g^2}{4\, F_\pi^2} 
\,\cdot \, \epsilon^2
\, , 
\ee
given by the decays 
$B_d^0\to \pi^+$, $B_s^0\to K^+$ and $B_u^-\to \pi^0,\eta_8,\eta_0$.   
Notice that at large $N_C$ the $\eta_0$ forms the ninth chiral
Goldstone boson.


The modifications to the  interaction vertices and couplings  
in the form factors (Eq.~\eqn{eq.FF}) are expected to be 
of the order of $\frac{1}{N_C}$  and $\frac{\Lambda_{\rm QCD}}{m_b}$.  

However, the gaining of mass by the $\eta_0$ meson is far more important.  
It generates large modifications in the available phase-space 
in Eq.~\eqn{eq.phasespace} and 
important kinematical corrections,  of the 
order of $\left(\frac{m_{\eta_0}}{\Lambda_E}\right)^2$.  
The chiral octet $(\pi,K,\eta_8)$ remains 
massless within  
the chiral limit whereas  the chiral singlet $\eta_0$ gains a mass 
$m_{\eta_0}\sim 900$~MeV due to $1/N_C$ corrections~\cite{meta0}. 
The available phase-space for the ${B_u^- \to \eta_0}$ mode gets reduced 
and the integration of the form factor (Eq.~\eqn{eq.FF}) 
over the available range of $z$ yields
\vspace*{-0.3cm}
\bel{eq.DG}
\ba{l}
\Delta \Gamma(\epsilon,r_\epsilon)_{B_u\to\eta_0}=
\Frac{1}{3}\,\times  \, \Frac{f_B^2 g^2}{4\, F_\pi^2} 
\times  \epsilon^2 
\\
\times\left[
\left(1+2 \re \right) \sqrt{1-\re} 
+\frac{3 \re}{2}\ln{\Frac{1+\sqrt{1-\re}}{1-\sqrt{1-\re}}}
\right]
\, .
\ea
\ee
The other decay modes remain unaffected.

\vspace*{-0.2cm}
\section{Results and conclusions}

By observing the $B^0 - B^-$ difference it is possible to separate 
the WA from other $1/m_b^3$ effects~\cite{Gambino}. 
The subtraction of the integrated widths 
$\Delta \Gamma(\epsilon)_{B_q\to X_u \ell \nu}$ 
gives
\vspace*{-0.3cm}
\bel{eq.WAdif}
\Frac{\left[ \Delta \Gamma_{B^-} -\Delta \Gamma_{B^0}\right]}{\Gamma_0} 
= 
\Frac{f_B^2 g^2}{2 F_\pi^2} \Frac{m_{\eta_0}^2}{M_B^2} 
%
%
\left\{1+\ln \Frac{\re}{4}\right\}  + ...
\ee
with  
the dots standing for contributions suppressed by
$\left(\Frac{m_{\eta_0}}{\Lambda_E}\right)^2$. 
This result reproduces all the features 
in the difference $[\Delta \Gamma_{B^-}-\Delta \Gamma_{B^0}]$ 
that one expects from the OPE:

\tab --Since $f_B^2\sim 1/M_B$, suppression by $1/M_B^3$.

\tab --Suppression 
by $1/N_C$: from the scaling 
$m_{\eta_0}^2\sim \frac{1}{N_C}$ with respect
to the other energy scales.  

\tab --The production of 
chiral singlets in the final state is responsible  of the difference
(WA~topologies in Fig.~\eqn{fig.WAdiagram}).

\tab --As $\Lambda_E$ grows,   the difference  
$\Delta \Gamma_{B^-}-\Delta \Gamma_{B^0}$ tends to a constant  
up to a residual term  
$\ln{\left(\frac{m_{\eta_0}}{\Lambda_E}\right)}$;    
the effect is non-perturbative and comes from the  end-point of the
$q^2$ spectrum.  
The removal of this residual log needs the consideration of
higher $B$ multiplets, in order to recover the $f_+\sim \frac{1}{E_P^2}$ 
behaviour of the form factor at high energies~\cite{LEET}. 
This have been implemented by
considering modified-pole expressions~\cite{modifiedpole}.

In Fig.~\eqn{fig.WAgraf}, we show the dependence of 
the difference of branching ratios 
$[\mB_{B_d^0}-\mB_{B_u^-}]$ on the cut-off, $\epsilon$. One can see how around
$\Lambda_E\sim $1~GeV, 
the difference stabilises and becomes approximately constant (upto the
residual log dependence referred to before), reaching the value 
\vspace*{-0.3cm}
\be
[\Delta\mB_{B_d^0}-\Delta\mB_{B_u^-}]
\sim 
4 \times \left(\Frac{f_B}{0.2\mbox{ GeV}}\,
\, \Frac{g}{0.5}\, |V_{ub}|\right)^2 \, , 
\ee
in agreement with former estimates~\cite{Voloshin,Gambino,Meyer}.

\begin{figure}[t!]
\begin{center}
\includegraphics[width=5.6cm,clip]{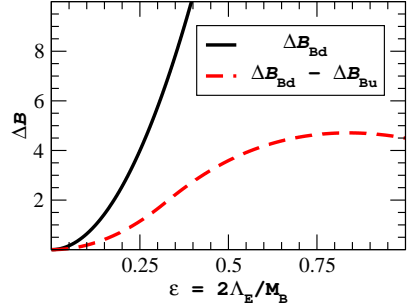}
\vspace*{-1cm}
\caption{
\small{
Dependence of the difference of branching ratios 
$[\Delta \mB_{B_d^0}-\Delta \mB_{B_u^-}]$ on the cut-off $\epsilon$ (dashed
curve).   
For comparison, it is shown together with $\Delta \mB_{B_d^0}$ (solid). 
Expressions are from  
Eqs.~\eqn{eq.LOFF}--\eqn{eq.WAdif} and 
$\Delta \mB$ is expressed in units of 
$\left(\frac{f_B}{0.2\mbox{ GeV}}\, \frac{g}{0.5}\, |V_{ub}|\right)^2$.
}}
\vspace*{-1.3cm}
\label{fig.WAgraf}
\end{center}
\end{figure}

This WA estimate shows how it is possible to
describe these OPE effects through HM$\chi$PT, 
recovering all the expected features.  

The next step is the calculation of the $B\to V \ell \nu$ contribution.  
The residual  $\ln{(r_\epsilon)}$ dependence in Eq.~\eqn{eq.WAdif} 
is expected to be removed 
through modified pole expressions~\cite{modifiedpole}.  
The relevance of this effect and the proposed techniques 
should be taken into account in the analyses of $D$ decays, 
where WA will be more important.

\end{document}